\pdfoutput=1
\documentclass[conference]{IEEEtran}
\IEEEoverridecommandlockouts
\usepackage{cite}
\usepackage{amsmath,amssymb,amsfonts}
\usepackage{algorithmic}
\usepackage{graphicx}
\usepackage{textcomp}
\usepackage[pscoord]{eso-pic}
\usepackage[a4paper, total={184mm,239mm}]{geometry}
\usepackage{comment}
\usepackage{xcolor}

\newcommand{\toolName}{\textit{BatchLens}}

\newcommand{\modify}[1]{\textcolor{black}{#1}}

\graphicspath{ {./figures/} }

\newcommand{\placetextbox}[3]{
  \setbox0=\hbox{#3}
  \AddToShipoutPictureFG*{
    \put(\LenToUnit{#1\paperwidth},\LenToUnit{#2\paperheight}){\vtop{{\null}\makebox[0pt][c]{#3}}}%
  }%
}%

\makeatletter
\newcommand{\linebreakand}{%
  \end{@IEEEauthorhalign}
  \hfill\mbox{}\par
  \mbox{}\hfill\begin{@IEEEauthorhalign}
}
\makeatother

\def\BibTeX{{\rm B\kern-.05em{\sc i\kern-.025em b}\kern-.08em
    T\kern-.1667em\lower.7ex\hbox{E}\kern-.125emX}}

\title{BatchLens: A Visualization Approach for Analyzing Batch Jobs in Cloud Systems}

\author{
\linebreakand
\IEEEauthorblockN{Shaolun Ruan}
\IEEEauthorblockA{School of Computing and Information Systems \\
\textit{Singapore Management University}\\
Singapore, Singapore \\
slruan.2021@phdcs.smu.edu.sg}
\and
\IEEEauthorblockN{Yong Wang}
\IEEEauthorblockA{School of Computing and Information Systems \\
\textit{Singapore Management University}\\
Singapore, Singapore \\
yongwang@smu.edu.sg}
\linebreakand
\IEEEauthorblockN{Hailong Jiang}
\IEEEauthorblockA{Department of Computer Science \\
\textit{Kent State University}\\
Kent, U.S. \\
hjiang13@kent.edu}
\and
\IEEEauthorblockN{Weijia Xu}
\IEEEauthorblockA{Scalable Computational Intelligence Group \\
\textit{Texas Advanced Computing Center}\\
Austin, U.S. \\
xwj@tacc.utexas.edu}
\and
\IEEEauthorblockN{Qiang Guan}
\IEEEauthorblockA{Department of Computer Science \\
\textit{Kent State University}\\
Kent, U.S. \\
qguan@kent.edu}

}
\placetextbox{0.5}{1}{\textit{\large This is a pre-print version. The final paper will be released by DATE 2022 when available.}}

\begin{document}

\maketitle

\begin{abstract}

Cloud systems are becoming increasingly powerful and complex. It is highly challenging to identify anomalous execution behaviors and pinpoint problems by examining the overwhelming intermediate results/states in complex application workflows. Domain scientists urgently need a friendly and functional interface to understand the quality of the computing services and the performance of their applications in real time. To meet these needs, 
we explore data generated by job schedulers and investigate general performance metrics (e.g., utilization of CPU, memory and disk I/O). 
Specifically, we propose an interactive visual analytics approach, \toolName{}, to provide 
both providers and users of cloud service
with an intuitive and effective way to explore the status of system batch jobs and help them conduct root-cause analysis of anomalous behaviors in batch jobs.
We demonstrate the effectiveness of \toolName{} through a case study on the public Alibaba bench workload trace datasets.



\end{abstract}


\section{Introduction}


Cloud computing has become the backbone of modern IT systems, which supports processing large volumetric data using clusters of computing nodes~\cite{b1,b2}. Understanding the batch jobs’ behaviors on cloud platforms is of great importance for cloud providers and cloud service users. Anomalous behaviors of batch jobs can potentially indicate existing software bugs and hardware crashes, which will eventually result in the violation of the Service Level Agreement (SLA)~\cite{b3}. However,
it is still a challenging and complex task to diagnose and prevent anomalous execution behaviors in cloud computing environments~\cite{b5}.

To prevent software flaws and hardware accidents
that can result in the failure of cloud services,
cloud providers have been monitoring cloud platforms through metrics-based~\cite{b6,b7}, log-based~\cite{b9} and trace-based~\cite{b11,b12} approaches. More recently, deep learning-based approaches 
are also used
for anomalous behavior detection~\cite{b14}. Prior studies, though, are effective techniques to monitor job and system behaviors, the cause is still invisible to the cloud system administrators due to the hidden patterns of the batch job co-allocation. Meanwhile, the existing tools are generally designed for system administrators, users may also need to monitor the status of their executing jobs so that they 
can provide more detailed information to system administrators when submitting the tickets. 
Moreover, the preceding methods are neither intuitive nor efficient as they consist of large-scale general metric data, which significantly affects the perception of the abnormal status of compute nodes and makes system administrators suffer from monitoring inflexibility.

\modify{Visualization tools have been extensively adopted to support offline log analysis in a variety of cloud applications.} Prior studies~\cite{b15} have demonstrated that visual representation can provide rich insights in monitoring cloud computing performance, and increase the possibility of uncovering hidden patterns in cloud infrastructures through intuitive visual representations and effective user interactions.

In this paper, we propose a visualization approach \toolName{} to analyze and monitor the job execution behaviors in cloud computing systems.
\modify{Compared with existing works, \toolName{} leverages effective visual representations and flexible interactions to analyze and detect anomalous batch jobs in cloud systems.}
Using the traces from large-scale parallel cloud systems at Alibaba, we develop multiple mutually-linked views to analyze the running jobs on metric-heavy compute nodes and enhance the effective human perception for the batch jobs. Specifically, interactive visual designs of hierarchical bubble charts and line charts are proposed to support analysis of abnormal jobs through temporal and spatial comparison.
The major contributions of this paper can be summarized as follows:

\begin{itemize}
    \item We propose a novel visualization approach \toolName{} based on batch hierarchy data to
    enable interactive analysis of
    the batch jobs in cloud systems.

    \item We conduct a case study on the public Alibaba trace datasets to demonstrate the effectiveness of our proposed approach.
\end{itemize}

\section{Dataset and DAG Batch Workloads}

Alibaba trace datasets~\cite{32} is part of the Alibaba Open Cluster Trace Program, which contains performance profile data collected from the Alibaba large-scale distributed cloud computing platform across 1300 machine batch jobs and a 24-hour duration. 
In this paper, we only focus on batch jobs and their dependencies. Each trace record in batch scheduler data includes the hierarchical structure for a compute node set at a 300-second resolution. 
For server usage data set, each row includes metadata of the node and the performance log of three metrics, i.e., GPU utilization, memory utilization and disk utilization at a one-second resolution.
A task has one or multiple instances running on the respective compute node. According to our data pre-processing, 75\% batch jobs contain only one task, while 94\% tasks have multiple instances. Note that each instance must be executed by only one compute node, and each compute node can run multiple instances simultaneously.

\section{Visual Design}


To analyze the batch job status
from the perspectives of both their spatial distribution and temporal evolution,
\toolName{} provides users with multiple linked views to reveal the insights of batch job scheduling (Fig. \ref{fig:teaser}).
Also, rich interactions are enabled to facilitate convenient explorations.

\subsection{Hierarchical Bubble Chart}

The hierarchical bubble chart is an overview of batch dependency, which provides a comprehensive hierarchy of batch jobs, tasks and instances (corresponding compute nodes). We adopt hierarchical bubble charts, as it can intuitively visualize the hierarchy of multiple nodes. 

\begin{figure}[t]
 \centering 
 \includegraphics[width=0.8\columnwidth]{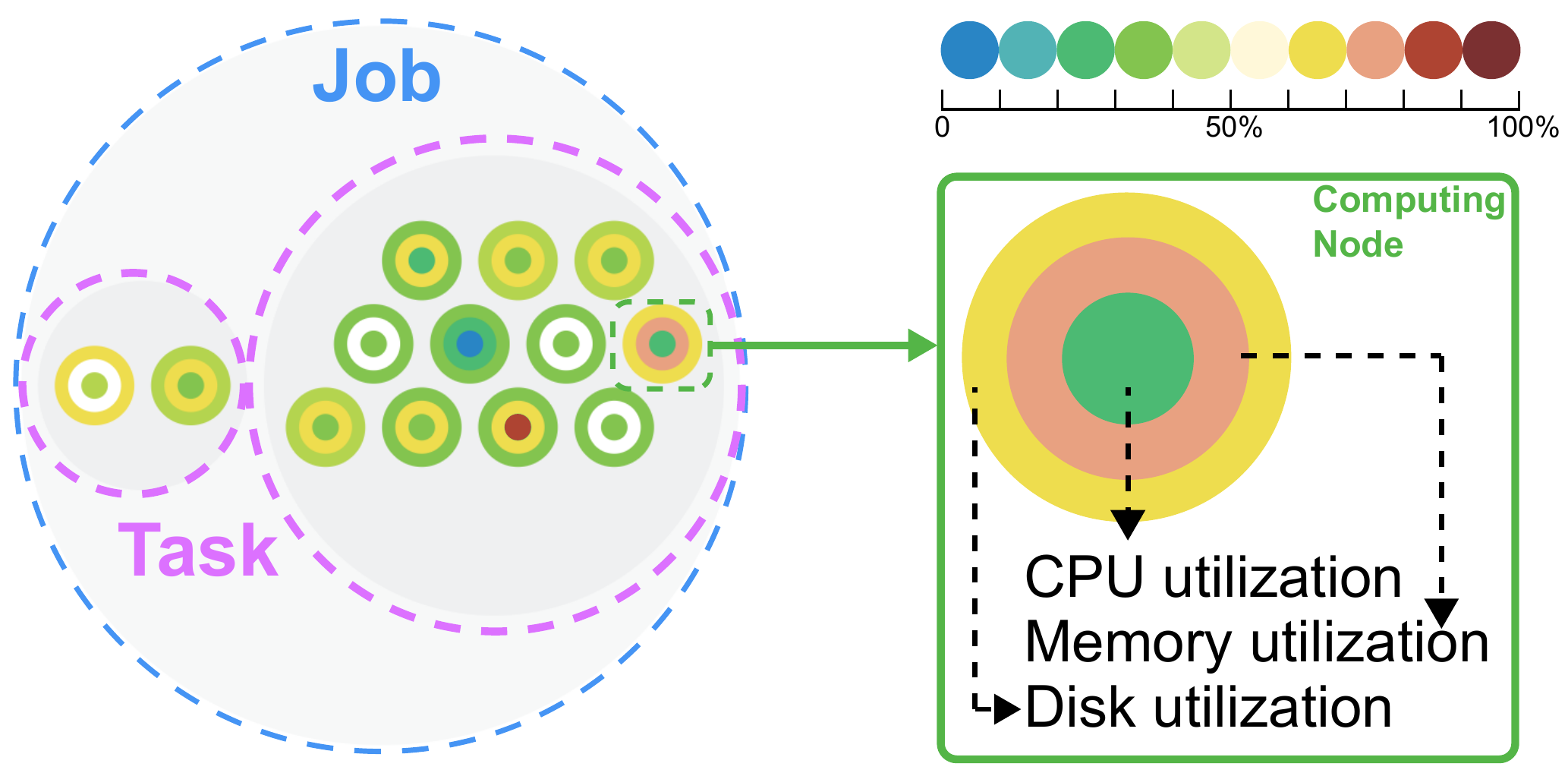}
 \caption{Visualization of batch scheduling data encoded by hierarchical bubbles and the color scheme for performance metrics, i.e., CPU utilization, memory utilization and disk utilization, which are indicated by three annuli in the detailed view.}
 \label{fig:2}
\end{figure}

Specifically, three layers of bubbles are applied to indicate the hierarchical batch entities (Fig. \ref{fig:2}). Bubbles highlighted with blue dotted circles denote batch job level, which contains the child level of tasks highlighted by purple dotted circles. Each compute node, which is scheduled to execute the batch instances and is subordinated to the respective task, is comprised of three parts denoting general usage metrics, i.e., CPU utilization, memory utilization and disk I/O utilization. We colorize the state metrics to reflect the performance of the machines at once.


\subsection{Line Charts}

Temporal analysis on cloud computing systems facilitates the detection of anomalous performances of compute nodes over time. We utilize line charts to reflect metric trends and incorporate multiple vertical annotation lines into line charts for start time and end time representation of batch jobs.

\begin{figure}[t]
 \centering 
 \includegraphics[width=\columnwidth]{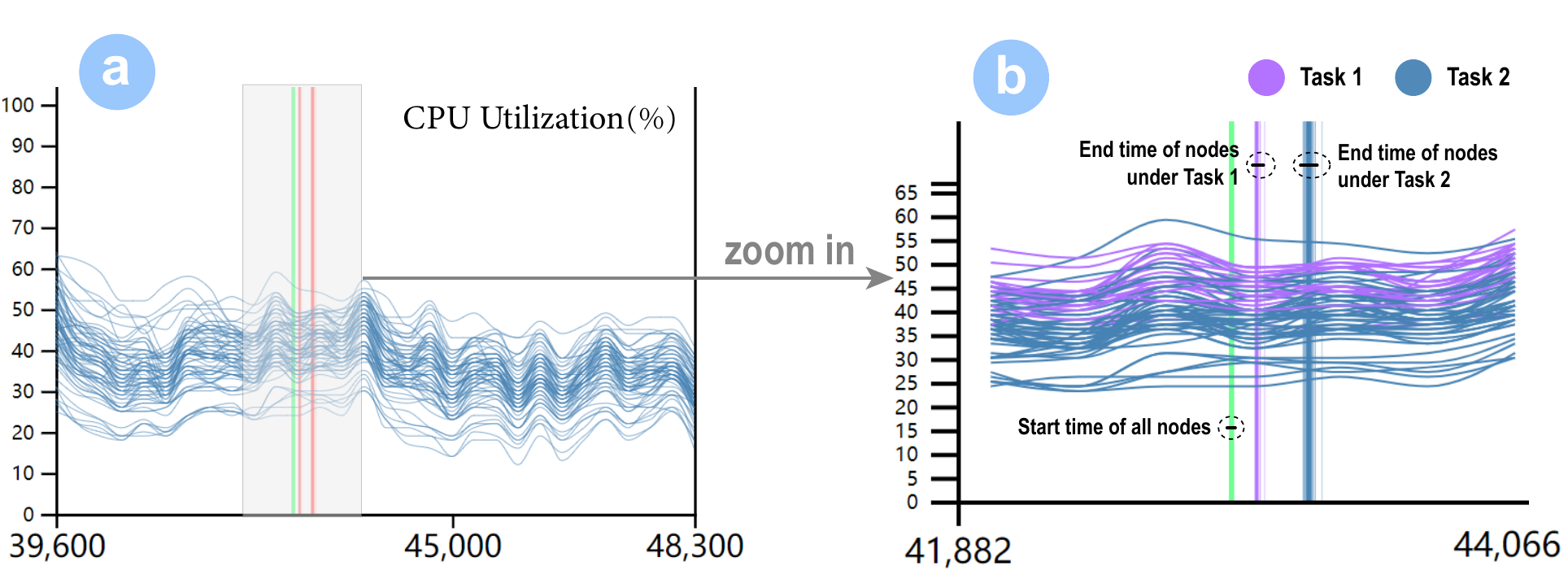}
 \caption{Multiple line chart reflecting the metric utilization changes of nodes under a selected batch job over time. (a) indicates the utilization trend in the overall time period. After selecting the time range via brushing, (b) is generated to show the detailed view of the selected part. For both views, vertical annotation lines in green and non-green are used to show the start time and end time of the job execution respectively.}
 \label{fig:3}
\end{figure}

As shown in Fig. \ref{fig:3}, line charts are used to indicate changes of machine utilization over time. Specifically, it shows metric trends of those compute nodes executing the same batch job simultaneously. For example, Fig. \ref{fig:3}(a) visualizes the CPU utilization of all the nodes executing \textit{job\_7399} in the overall time period. Green annotation lines denote the start time of job execution on corresponding nodes. All lines bundling into one cluster indicates that the job is scheduled for all nodes at the same time. Red lines depict end timestamps of job execution, which are bundled as two clusters, 
as \textit{job\_7339} includes two tasks and each has a different end timestamp. Also, after selecting the interesting time range of overall line charts by brushing, users can explore the detailed metric utilization (Fig. \ref{fig:3}(b)). In the selected detailed view, different lines and annotation lines are plotted in various colors, which enables users to compare node usages by task dimension. 
By interacting with the line charts of various batch jobs, users can observe the temporal patterns in terms of metric trends of compute nodes, such as a spike or a valley in the context of other nodes' performance.

\subsection{User Interactions}

Flexible and intuitive interactions allow a comprehensive analysis of the batch job observation. A simple timeline is used to represent the metrics aggregated across the entire cloud systems over time. Each layer of the graph represents one metric. Users can select an interesting time range and timestamp through the brushing and choosing interaction respectively.

\begin{figure*}[t]
 \centering 
 \includegraphics[width=\linewidth]{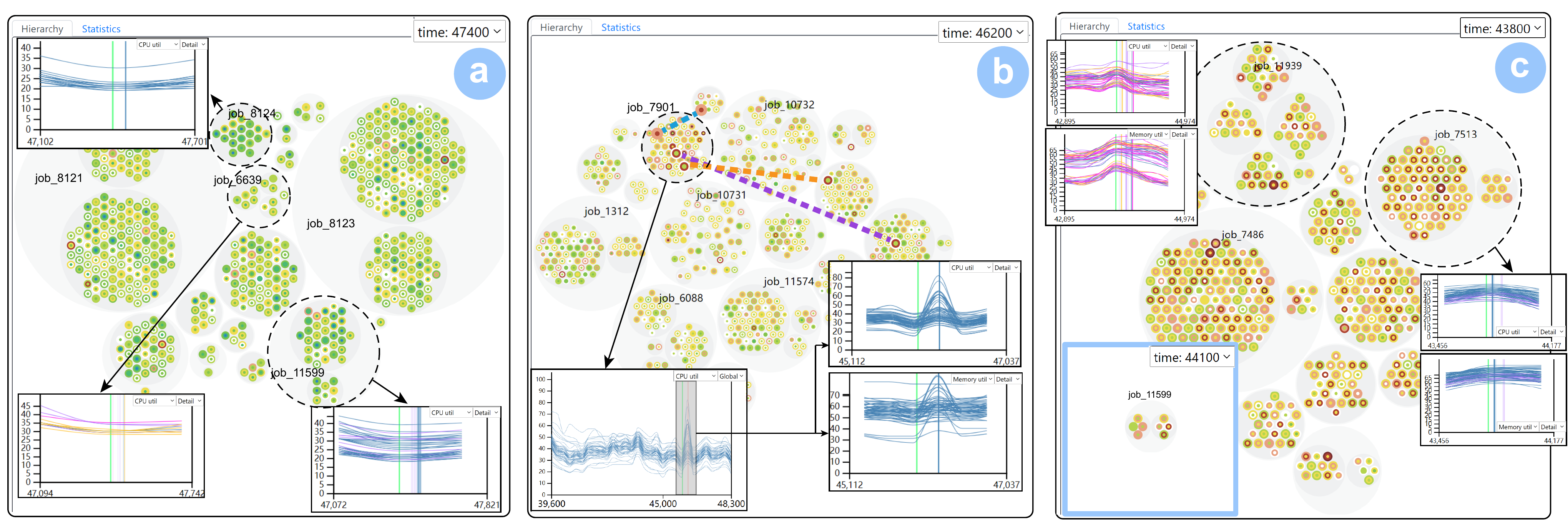}
 \caption{The main view and corresponding detailed views of the visual analytics system. Detailed views of multiple line charts show the metric utilization trends of compute nodes under a selected batch job over time. The green annotation lines in each line chart depict the start timestamps of the selected job across each compute node executing it, while other annotation lines denote the job end timestamps using the same color as the corresponding lines of nodes.}
 \label{fig:teaser}
\end{figure*}

Also, another simple user interaction is adopted to recognize the same compute node executing various batch jobs simultaneously. Since the hierarchical bubble chart is a job-based graph, the same node may be rendered into multiple parent job bubbles. A direct mouse over on target nodes will trigger a zoom-in refresh, as shown in pairwise nodes that are linked with the same color dotted lines in Fig. \ref{fig:teaser}(b).


\section{case study}

We conduct a case study using Alibaba trace datasets to demonstrate the effectiveness of the proposed visualization approach. We select three interesting and representative timestamps to illustrate the hidden patterns of abnormal machine status (Fig. \ref{fig:teaser}). It is clear that both figures are uniform in color distribution due to the load balance.

Fig. \ref{fig:teaser}(a) shows a common situation of the system, where all the machines that host some tasks are at low resource utilization (20\% - 40\%), and all the performance metrics are stable. Specifically, from the hierarchical bubble chart, we can observe that there are 15 root bubbles denoting batch jobs at timestamp 47400, which includes two primary jobs (Job \textit{job\_8121} and Job \textit{job\_8123}), both of which are scheduled into two tasks that are executed a substantial volume of nodes. More specifically, Job \textit{job\_8124}, which is scheduled into one task only, has the nodes with the lowest utilization (CPU, memory and disk I/O). From the line chart corresponding to Job \textit{job\_8124}, we can see that the CPU utilization of all nodes is fairly constant with only small increase during the period of job execution ( between the creation and termination of job). Also, for Job \textit{job\_6639}, the lines denoting CPU utilization of nodes under four separated tasks are plotted with four colors. 
These four types of tasks, as the annotation lines plots, have the same start timestamp but multiple end timestamps. However, CPU utilization for all the nodes in different tasks stays stable during job executions. The same pattern can be observed for Job \textit{job\_11599}.

In bubble chart of Fig. \ref{fig:teaser}(b),
it is clear that all nodes are running at medium level of resource utilization around Timestamp 46200 (50\%-80\%). The resource utilization on the nodes hosting the jobs  are heavier than that in Fig. \ref{fig:teaser}(a) through the color distribution of bubble charts, with an exception of Job \textit{job\_7901} running on busier nodes than those hosting other jobs. 
From the line chart of the overall time period of job of \textit{job\_7901} (left bottom view), we can see that the CPU utilization of corresponding nodes is synchronized, even though drastic fluctuations exist. From the detailed view (right bottom view), a notable spike emerges for CPU and memory usage after Job \textit{job\_7901} is scheduled into the corresponding machines. Both metrics reach the peak of the utilization when the job execution is over, followed by a slow drop to the normal level, which indicates that the machines running Job \textit{job\_7901} experience intensive workload during the execution time. Additionally, we connect the same machines with colored dotted lines (green, orange and purple) in the bubble chart to help trace down the machines execute multiple tasks simultaneously.


More interesting findings can be revealed from Fig. \ref{fig:teaser}(c). A tremendous amount of nodes are running at high CPU- or memory-utilization at Timestamp 43800, including several nodes reaching the respective capacity of node performance. 
From the line charts generated from Job \textit{job\_7513} at the bottom right, CPU utilization of nodes under two tasks is distinguished by blue and purple lines. Line cluster in purple depicts the relatively smaller task set, which has a less severe CPU and memory usage.
Also, for Job \textit{job\_11939}, lines denoting five different tasks are entangled seriously as shown in the detailed view on the top left. 
The same pattern can be perceived between views for CPU and memory utilization: an obvious drop occurs after the creation of Job \textit{job\_11939}. Moreover, we find that at Timestamp 44100, all of the preceding nodes on the system are shut down (figure at bottom left in Fig. \ref{fig:teaser}(c)), and only Job \textit{job\_11599} is left on the entire platform. However, the general metrics still exist for the corresponding machines at Timestamp 44100 in the detailed line chart on the top left. It is likely to speculate that the compute node is suffering thrashing while the virtual memory is overused with the degree from multi-programming increasing. 
Eventually thrashing forces the CPU utilization to decrease and the whole system is not making any progresses. From the observation in the next time slice when almost all jobs disappear, these jobs are very likely terminated and relaunched by the user or system administrator to clear the thrashing. 


\section{Related Work}

This work is related to prior research on anomalous behavior detection in cloud systems and visualization for cloud computing analysis.

\subsection{Anomalous Behavior Detection in Cloud Systems}

Anomalous behavior detection of the distributed system is an essential and challenging topic which attracts great research attention. Prior works can be categorized into three groups, i.e., metric-based, log-based and trace-based approaches. Metric-based~\cite{19} approach usually applies statistics on collected metrics, which mainly include performance metrics, e.g., throughput and system metrics, e.g., CPU, memory and I/O. 
Logs are generated along with applications’ running and reflect application processing status and execution logic. Recent studies~\cite{23,24} leverage logs or traces to create workflow models for software testing and the understanding of system behaviors. Lou et al.~\cite{22} proposed an automaton model for reconstructing concurrent workflows from event traces. Traces record information for program debugging or diagnosis purposes. The preceding studies are not preferred as without specifically-designed visualization methods, inflexible row cloud trace data can not be presented by intuitive visual summarization providing quick and efficient analytic process.

\subsection{Visualization for Cloud Computing Analysis}

Existing visual metaphors~\cite{add0, add1} have studied representations on comparison of quantitative state changes (e.g., hardware metrics of compute nodes), while they are not suitable for our application scenario as the trace data includes spatial characteristics for topological batch distribution.
Many visualization tools~\cite{28,29} serve as collecting low-level trace data from large parallel systems. More recently, Muelder et al.~\cite{b16} proposed a typical system for cloud computing analysis to portray the behavior of each compute node over time. A variety of visualizations~\cite{30} have been proposed for monitoring and analyzing trace data generated from cloud computing systems. \modify{Though the preceding studies provide rich insights of large-scale parallel network analysis, they rarely analyze the anomalous batch jobs, which supports the anomalous behavior detection and further conducts root-cause analysis.}

\section{Conclusion and Future Work}

We propose a visualization approach to interactively analyze the execution behavior using general performance metrics from Alibaba trace datasets. We have demonstrated the effectiveness with case studies and revealed three existing patterns of batch jobs on cloud systems, all of which can support the perception of the hidden reasons behind hardware-heavy compute nodes. 
Although our technique may not present every facet of the reasons for the anomalies, 
it provides system administrators and cloud users with deep insights into batch job status, and facilitate an easy detection of anomalies.


In future work, we would like to further extend our approach from two directions.
First, real-time techniques have been extensively adopted on large-scale cloud computing platforms.
We plan to extend \toolName{} into a real-time online system and integrate it into real cloud distributed systems. 
Furthermore, 
\toolName{} is an effective approach to detect abnormal jobs through visualizing their hardware performance metrics. But some hidden anomalies will not affect hardware performance significantly due to the load balance.
It will be interesting to further investigate how to recognize those hidden abnormal statuses.

\end{document}